\documentstyle[12pt,epsf]{article}

\renewcommand{\thefootnote}{\fnsymbol{footnote}}
\newcommand{\EQ}{\begin{equation}}
\newcommand{\EN}{\end{equation}}
\newcommand{\bea}{\begin{eqnarray}}
\newcommand{\ena}{\end{eqnarray}}
\newcommand{\bear}{\begin{array}}
\newcommand{\enar}{\end{array}}
\newcommand{\iran}{\nonumber}
\newcommand{\vs}[1]{\vspace{#1 mm}}

\newcommand{\nami}{\widetilde}

\newcommand{\uda}{\nearrow \kern-1em \searrow}

\newcommand{\NP}[1]{Nucl.\ Phys.\ {\bf #1}}
\newcommand{\PL}[1]{Phys.\ Lett.\ {\bf #1}}

\newcommand{\PR}[1]{Phys.\ Rev.\ {\bf #1}}
\newcommand{\PRL}[1]{Phys.\ Rev.\ Lett.\ {\bf #1}}
\newcommand{\PTP}[1]{Prog.\ Theor.\ Phys.\ {\bf #1}}

\begin{document}

\topmargin 0pt
\oddsidemargin 5mm

\begin{titlepage}
\setcounter{page}{0}
\begin{flushright}
OU-HET 278 \\
hep-ph/9709367 \\
September, 1997
\end{flushright}

\vs{12}
\begin{center}
{\Large Reconstruction of quark mass matrices in the NNI form \\
  from the experimental data}
\vs{15}

{\large 
Eiichi Takasugi\footnote{
               e-mail address: takasugi@phys.wani.osaka-u.ac.jp}
and Masaki Yoshimura\footnote{
                    e-mail address: masaki@phys.wani.osaka-u.ac.jp}\\}
\vs{8}
{\em Department of Physics, Osaka University \\ 
  1-16 Toyonaka, Osaka 560, Japan} \\
\end{center}
\vs{6}

\centerline{{\bf{Abstract}}}

We examined the question that what is a general form of quark mass 
matrices which is achieved by the transformation that leaves the 
left-handed gauge interaction invariant. In particular, we analyzed 
in detail 
the Fritzsch-type and the Branco-Silva-Marcos-type parametrization. 
Both parametrizations contain ten parameters and can be expressed by 
the experimental data. We explicitly reconstructed quark mass matrices 
in terms of quark masses and CKM parameters for the Fritzsch-type 
parametrization.  

\end{titlepage}
\newpage
\renewcommand{\thefootnote}{\arabic{footnote}}
\setcounter{footnote}{0}  

\section{Introduction} \label{sec:Intro}

There are increasing interests in understanding the CKM mixings 
and the CP violation parameter. Many works were made to find 
$Ans\ddot{a}tze$ for quark mass matrices and discussed its predictions. 
The Fritzsch $Ansatz$[1] is one of examples. 
This approach is  called  the texture zero analysis where 
some elements of mass matrices are required to be zero to reduce 
the degrees of freedom in mass matrices. The extensive 
works are made along this line[2]-[5]. 

Recently, some attention was paid to the inverse problem, that is, 
to reconstruct quark mass matrices directly from the experimental data. 
This problem is ambiguous until the form of quark mass matrices is 
fixed. Therefore, we need to choose a general and convenient 
form of quark mass matrices to start with. One of candidates is 
the nearest neighbor interaction 
(NNI) form which is considered as a ``general" form of quark 
mass matrices because this form is achieved by the 
transformation that leaves the left-handed gauge interaction invariant[6]. 
Harayama and Okamura[7] chose the NNI form and showed directly how 
elements of quark mass 
matrices are expressed by observables, quark masses and CKM parameters. 
However, the NNI form of 
quark mass matrices contain twelve parameters so that all parameters 
are not fixed uniquely because there are  only ten observables 
for three generation case. Hereafter, we consider only three generation 
case. Subsequently, Koide[8] showed that within the NNI form   
the 3-2 element of up-quark mass matrix can be made zero  
in general. This is achieved by using the 
rephasing freedom of quarks. Now the number of parameters in this 
parametrization is ten so that quark mass matrices are completely 
determined by the data. 

Recently, one of authors showed[9] that  quark 
mass matrices  can be transformed in general to either one of 
the following two forms,  the Fritzsch-type parametrization 
or  the Branco-Silva-Marcos(BS)-type parametrization. 
The Fritzsch-type parametrization is that the u-quark mass matrix 
is in a Fritzsch form[1], while the d-quark mass matrix is in the 
NNI form. The BS-type parametrization is  that  
 the d-quark mass matrix is in  the
Branco-Silva-Marcos form[10], while the u-quark mass matrix is in 
the NNI form. The transformation to these forms is constructed 
explicitly in Ref.[9] and examined for what quark mass matrices 
the transformation is possible. 
In both parametrizations, quark mass matrices 
contain ten physical parameters so that they can be determined 
uniquely by  quark masses and the  CKM parameters. The formula 
how to express quark mass matrices in terms of data are given. 
Falcone, Pisanti and Rosa[11] reconstructed quark mass matrices 
by using some special form:  $M_u$ is in a diagonal form and $M_d$ 
has zero at the 1-1, 2-2 and 3-1 elements, and the form where 
$M_u$ and $M_d$ are exchanged.   

In this paper, we analyze the transformation in detail and 
examine for what form of quark mass matrices  
the transformation to the Fritzsch-type or the BS-type parametrization 
is possible. Our answer is that the transformation is possible for 
most of physically meaningful quark mass matrices, that is, those 
which reproduce the present data of quark masses and CKM mixing 
parameters well. Then we reconstruct quark mass matrices from the 
experimental data in the Fritzsch-type parametrization. 

In below, we shall briefly explain the transformation as an introduction. 
First, we give the transformation of quark mass matrices to the 
Fritzsch form: 
\EQ
 U^{\dagger}M_{u}V_{u}=\nami{M}_{uF}\;,
 \quad U^{\dagger}M_{d}V_{d}=\nami{M}_{d}\;,
\EN
where 
\EQ
 \nami{M}_{uF}=\pmatrix{0 & a_{u} & 0 \cr 
                 a_{u} & 0 & c_{u} \cr 
                 0 & c_{u} & e_{u}}\;,
\EN
and
\EQ
 \nami{M}_{d}=\pmatrix{0 & a_{d}e^{i \alpha_{1}} & 0 \cr 
                        b_{d} & 0 & c_{d}e^{i \alpha_{2}} \cr 
                        0 & d_{d} & e_{d}}\;,
 \label{mat:dNNI-form}
\EN
under the condition that the equation for the u-quark phases $\theta_j$,  
\EQ
 \sum_{j,k}e^{i(\theta_{j}-\theta_{k})}(O_{u})_{1j}(O_{u})_{2k}
  (KD_{d}^2K^{\dagger})_{jk}=0\;,
\EN
has the solution. This equation contains two phases 
$\theta_{12}\equiv \theta_{1}-\theta_{2}$ and 
$\theta_{23}\equiv\theta_{2}-\theta_{3}$ which are to be fixed by 
this equation. 
Here, $D_d$ is the diagonal mass matrix for d-quarks,  
$K$ is the CKM matrix and $O_u$ is the orthogonal matrix which 
diagonalizes the Fritzsch matrix $\nami{M}_{uF}$. 
Elements of u-quark matrix $ \nami{M}_{uF}$ are written explicitly by
the u-quark masses and d-quark mass matrix $\nami{M}_{d}$ are given by 
\bea
  a_{d}&=&\sqrt{H_{d11}}\;, \qquad 
  b_{d}=\sqrt{ \frac{ \det{H_{d}} }
             {H_{d11}H_{d33}-|H_{d13}|^2} }\;, \iran\\
  c_{d}&=&\frac{ |H_{d23}| \sqrt{H_{d11}} }
         {\sqrt{ H_{d11}H_{d33}-|H_{d13}|^2 }}\;, \quad
  d_{d}=\frac{ |H_{d13}| }{ \sqrt{H_{d11}} }\;, \iran\\
  e_{d}&=&\frac{ \sqrt{ H_{d11}H_{d33}-|H_{d13}|^2} }
         { \sqrt{H_{d11}} }\;, \label{gen:F-NNI} \\
  \alpha_{1}&=&\arg{ H_{d13} }\;, \qquad 
  \alpha_{2}=\arg{ H_{d23} }\;, \iran
\ena
where elements of the matrix $H_d$ are given by 
\EQ
 (H_{d})_{il}= 
  \sum_{j,k}e^{i(\theta_{j}-\theta_{k})}(O_{u})_{ij}(O_{u})_{lk}
  (KD_{d}^2K^{\dagger})_{jk}\;.
\EN
The matrix $H_d$ contains u-quark phases which are determined 
from Eq.(4) and all others are determined by quark masses 
and the CKM matrix. 

Similarly, the transformation of quark mass matrices to the BS form is 
obtained. Here, $M_d$ is transformed to the BS form
\EQ
 \nami{M}_{dBS}=\pmatrix{0 & a_{d} & 0 \cr 
                 a_{d} & 0 & c_{d} \cr 
                 0 & e_{d} & e_{d}}\;,
\EN
while all others are obtained 
by just changing the suffix u to d and the phase 
$\theta_j$ to $\phi_j$ and $K$ to $K^\dagger$. Thus the condition 
for the existence of the transformation becomes 
\EQ
 \sum_{j,k}e^{i(\phi_{j}-\phi_{k})}(O_{d})_{1j}(O_{d})_{2k}
  (K^{\dagger}D_{u}^2K)_{jk}=0\;.
 \label{cond:NNI-BS}
\EN
Here, $D_{u}$ is the diagonal mass matrix for u-quarks and $O_{d}$ is
the orthogonal matrix which diagonalizes
$\nami{M}_{dBS}\nami{M}_{dBS}^{\dagger}$.

In the following, we solve equations in Eq.(4) and (8) and examine 
in what region of CKM parameters the solution exists  and 
reconstruct the quark mass matrices from the experimental data 
for the Fritzsch-type parametrization. 
 
In Sec.2, we see how $M_{u}$ can be transformed to a Fritzsch form or
how $M_{d}$ can be transformed to a BS form. 
We obtain a complex equation containing two phases for each case.
In Sec.3, we show that these equations have solutions for two phases,
when mass matrices are ones which reproduce the CKM quark mixings.
Summary is given in Sec.4.

\section{Region of CKM parameters where 
the transformation exists}
We examine the equation for quark phases given in Eq.(4) or (8). 
If the solution exists, the transformation of quark mass matrices 
to the Fritzsch-type form or the BS-type form becomes possible. 
It is hard to solve these equation explicitly so that we try to 
solve perturbatively with respect to a small parameter, Cabibbo 
angle $\lambda$. For this purpose, we first  parametrize the CKM 
matrix in a Wolfenstein form 
\bea
 K \simeq \pmatrix{
   1-\frac{\lambda^2}{2} & \lambda &  \lambda^4 \sigma e^{i \delta}\cr
     -\lambda & 1-\frac{\lambda^2}{2} & \lambda^2 A \cr
     \lambda^3 \sigma'e^{i \delta'} & -\lambda^2 A & 1 }\;,
\ena
where $\sigma$, $\sigma'$, $\delta$ and $\delta'$ are related to 
$\rho$ and $\eta$ in the original Wolfenstein parameter as 
\EQ
 A\rho=\lambda
     \sigma\cos{\delta}=A-\sigma'\cos{\delta'}\;, \quad
   A\eta=\lambda 
     \sigma\sin{\delta}=\sigma'\sin{\delta'}\;.
\EN
We define ratios of quark masses as
\bea
 \left.
  \bear{ll}
   r_{u} \equiv (m_{u}/m_{c})/\lambda^4 \sim O(1)\;,& \quad
   r_{c} \equiv (m_{c}/m_{t})/\lambda^4 \sim O(1)\;,\\
   r_{d} \equiv (m_{d}/m_{s})/\lambda^2 \sim O(1)\;,& \quad
   r_{s} \equiv (m_{s}/m_{b})/\lambda^{5/2} \sim O(1)\;.
  \enar
 \right.
\ena

When we need to estimate numerically, we use 
$\lambda=0.2205$ for the Cabibbo angle 
and the running quark masses in units of GeV defined
at $\mu=M_{Z}$~\cite{FK},
\bea
 \left.
  \bear{lll}
    m_{u}=0.00222\;,& \quad m_{c}=0.661\;,& \quad m_{t}=180\;,\\
    m_{d}=0.00442\;,& \quad m_{s}=0.0847\;,& \quad m_{b}=2.996\;.
  \enar
 \right.
 \label{ryo:quark-mass}
\ena

 \subsection{The case of the Fritzsch-type parametrization}
In order to solve the equation (4) for phases, $\theta_{12}=\theta_1-
\theta_2$ and $\theta_{23}=\theta_2-\theta_3$, we need the orthogonal 
matrix $O_u$ which diagonalizes the Fritzsch mass matrix for u-quark and 
$KD_d^2K^\dagger$. The matrix $O_u$ is expressed in terms of 
ratios of up-quark masses as 
\EQ
 O_{u} \simeq \pmatrix{
       1 & -\lambda^2 \sqrt{r_{u}} 
         & \lambda^8 r_{c} \sqrt{r_{u}r_{c}} \cr
       \lambda^2 \sqrt{r_{u}} & 1 
         & \lambda^2 \sqrt{r_{c}} \cr
       -\lambda^4 \sqrt{r_{u}r_{c}} 
         & -\lambda^2 \sqrt{r_{c}} & 1}\;.
\EN
Next, the matrix $KD_{d}^2K^{\dagger}$ is estimated by using  
d-quark masses and CKM parameters.  In the leading order of $\lambda$, 
we find 
\bea
 \left.
  \bear{ll}
   (KD_{d}^2K^{\dagger})_{11} \simeq m_{b}^2 \lambda^7 
                       (r_{s}^2 + \lambda\sigma^2)\;,& \quad
   (KD_{d}^2K^{\dagger})_{12} \simeq m_{b}^2 \lambda^6 
                       (A \sigma e^{-i\delta}+ r_{s}^2)\;,\\
   (KD_{d}^2K^{\dagger})_{22} \simeq m_{b}^2 \lambda^4 
                       (A^2 + \lambda r_{s}^2)\;,& \quad 
   (KD_{d}^2K^{\dagger})_{13} \simeq m_{b}^2 \lambda^4 \sigma 
                       e^{-i\delta}\;,\\
   (KD_{d}^2K^{\dagger})_{33} \simeq m_{b}^2\;,& \quad
   (KD_{d}^2K^{\dagger})_{23} \simeq m_{b}^2 \lambda^2 A\;.
  \enar
 \right.
\ena

By keeping leading order terms of $\lambda$, the equation (4) 
is expressed as
\EQ
 A_{1} e^{i \theta_{12}} - B_{1} e^{i \theta_{23}}+
 C_{1} e^{i(\theta_{12}+\theta_{23}-\delta)} = D_{1}\;,
 \label{kaku:theta12}
\EN
where
\bea
 \left.
  \bear{ll}
   A_{1}=\lambda^6 (A \sigma e^{-i \delta}+r_{s}^2)
    \equiv |A_{1}|e^{-i \kappa}\;,& \quad
   B_{1}=\lambda^6 A \sqrt{r_{u}r_{c}}\;,\\
   C_{1}=\lambda^6 \sigma \sqrt{r_{c}}\;,& \quad
   D_{1}=\lambda^6 \sqrt{r_{u}} (A^2 + \lambda r_{s}^2)\;.
  \enar
 \right.
\ena

This is a complex valued equation with two variables $\theta_{12}$ and 
$\theta_{23}$. From the equation
 $
 |\exp({i(\theta_{12}-\delta)})/
         (|A_{1}|\exp({-i(\kappa-\delta)})+C_{1}\exp({i\theta_{23}}))| 
       =|( D_{1} +B_{1} e^{i \theta_{23}} )|$, we find 
\bea
 \cos{(\theta_{23}-\zeta)}
     &=&\frac{ |A_{1}|^2 +C_{1}^2 -B_{1}^2 -D_{1}^2 }{2N}
            \equiv \frac{Z}{2N}\;, 
\ena
where,
\bea
    N& =& \sqrt{ p^2 + q^2 + 2pq \cos{\delta} }\;,\nonumber\\
    \cos{\zeta}& =& -(p + q\cos{\delta} )/N\;,\\
    \sin{\zeta}& =& -(q \sin{\delta})/N\;,\nonumber
\ena
with
\bea 
 \left.
  \bear{l}
    p =\lambda^{12} A \sqrt{r_{c}} [\sigma^2 -r_{u} (A^2 +\lambda
           r_{s}^2) ]\;,\\
    q = \lambda^{12}\sigma r_{s}^2 \sqrt{r_{c}}\;.
  \enar
 \right.     
\ena
The condition of the existence of the solution of $\theta_{23}$ 
and also  $\theta_{12}$ is $-1\le Z/2N \le 1$. 
 By using quark masses in Eq.(12) and 
 $\lambda$, this condition gives the   region in the $\rho-\eta$ 
 plain for a given $A$.  
In Fig.1 we show this region  by taking into account one standard 
deviation of the experimental value $A$. 
In these figures, a region surrounded by the various circles is 
an experimentally allowed area. The dotted area is the region where 
the solution exists, that is, the transformation to the Fritzsch-type 
parametrization is possible. This region is almost independent 
of the change of $A$. From these figures, we can say that the 
transformation is possible for most physically allowed region except 
for a small area in the first quadrant of CP phase. 
The vertex point of the triangle corresponds to the Fritzsch-BS $Ansatz$
case where a Fritzsch form is assumed for the up-quark mass 
matrix  and the BS form for the d-quark one. This $Ansatz$ contains 
eight free parameters so that the position in the $\rho-\eta$ plane 
is fixed~\cite{Ito}. 

 \subsection{The case of the BS-type parametrization}
 
Analysis can be made similarly to the Fritzsch case. 
The matrix $O_d$ which diagonalizes the BS mass matrix for d-quarks 
is given by
\EQ
 O_{d} \simeq \pmatrix{
        1 & -2^{-1/4}\lambda\sqrt{r_{d}} 
          & 2^{-1/4}\lambda^{7/2} r_{s}\sqrt{r_{d}} \cr
        2^{-1/4}\lambda\sqrt{r_{d}} & 1 
          & \lambda^{5/2} r_{s} \cr
        -2^{3/4}\lambda^{7/2}r_{s}\sqrt{r_{d}}
          & -\lambda^{5/2} r_{s} & 1
        }\;
\EN
and $K^\dagger D_u^2K$ is given by 
\bea
 \left.
  \bear{ll}
   (K^{\dagger}D_{u}^2K)_{11} \simeq 
                    m_{t}^2 \lambda^6 \sigma'^2\;,&\quad
   (K^{\dagger}D_{u}^2K)_{12} \simeq 
                   -m_{t}^2 \lambda^5 A \sigma'e^{-i\delta'}\;,\\  
   (K^{\dagger}D_{u}^2K)_{22} \simeq m_{t}^2 \lambda^4 A^2\;,&\quad
   (K^{\dagger}D_{u}^2K)_{13} \simeq 
                    m_{t}^2 \lambda^3 \sigma'e^{-i\delta'}\;,\\
   (K^{\dagger}D_{u}^2K)_{33} \simeq m_{t}^2\;,&\quad
   (K^{\dagger}D_{u}^2K)_{23} \simeq -m_{t}^2 \lambda^2 A\;.
  \enar
 \right.
\ena

By keeping leading order terms, the equation (8) is expressed as
\EQ
 A_{2} e^{i(\phi_{12}-\delta')}- iB_{2}\sin{\phi_{23}}-
 C_{2} e^{i(\phi_{12}-\delta'+\phi_{23})} = -D_{2}\;,
\EN
where
$\phi_{12}=\phi_{1}-\phi_{2}, \phi_{23}=\phi_{2}-\phi_{3}$,
and 
\bea
 \left.
  \bear{ll}
   A_{2}=\lambda^5 A \sigma'\;,& \quad
   B_{2}=2^{3/4}\lambda^{11/2} A r_{s} \sqrt{r_{d}}\;,\\
   C_{2}=\lambda^{11/2} \sigma' r_{s}\;,& \quad
   D_{2}=2^{-1/4}\lambda^5 \sqrt{r_{d}} (A^2-\lambda r_{s}^2)\;.
  \enar
 \right.
\ena
We find  a solution for $\phi_{23}$ as 
\bea
 \cos{\phi_{23}}=
   \lambda^{-1/2}\left[ \sqrt{2} \sigma'^2/r_{d} 
               -(A^2 +\lambda r_{s}^2) \right]/2 A r_{s}\;.
\ena
There is another solution, 
$\lambda^{-1/2}(A^2 +\lambda r_{s}^2)/2 A r_{s}$, but this is 
unphysical because it is larger than one if we use the current 
CKM data and quark masses given in Eq.(12).
Once $\phi_{23}$ has a solution, $\phi_{12}$ also has a solution. 
We show the result in Fig.2, where the dotted area is an allowed region 
 similarly to the Fritzsch-type parametrization case.   
This region is almost independent 
of the change of $A$ and extends all experimentally allowed area. 
So we conclude that the transformation to the BS-type parametrization 
is possible for all allowed  parameters region. The point indicated by 
F-BS type corresponds to the case where $M_u$ takes a Fritzsch form and 
$M_d$ does a BS form where unique prediction for $\rho$ and $\eta$ is 
obtained.  
\section{Reconstruction of the quark mass matrices for the 
 Fritzsch-type parametrization}

The u-quark mass matrix is in the Fritzsch form so that it is 
expressed by u-quark masses so that we concentrate $\nami M_d$. 
This can be made  by using a formula given in Eq.(5). For this, 
we have to evaluate $H_d$ first and the angles $\theta_{12}$ and 
$\theta_{23}$ next.  By using the expressions in (13) and (14), 
we find 
\bea
 H_{d11} &\simeq& m_{b}^2 \lambda^7
         \left[ r_{s}^2 + 
           \lambda \left\{ \sigma^2 +A^2 r_{u} -2\sqrt{r_{u}} 
                       ( A \sigma \cos{(\theta_{12}-\delta)}
                        +r_{s}^2 \cos{\theta_{12}}) 
                   \right\} 
         \right]\;, \iran\\
 H_{d22} &\simeq& m_{b}^2 \lambda^4
         \left[ A^2 + r_{c} + 2 A \sqrt{r_{c}} 
            \cos{\theta_{23}} +\lambda r_{s}^2 \right]\;,\iran\\
 H_{d33} &\simeq& m_{b}^2\;, \label{ele:Hd}\\
 H_{d13} &\simeq& m_{b}^2 \lambda^4
         \left[ \sigma e^{i(\theta_{12}-\delta+\theta_{23})}
               -A \sqrt{r_{u}} e^{i\theta_{23}} 
         \right]\;,\iran\\
 H_{d23} &\simeq& m_{b}^2 \lambda^2
         \left[ A e^{i\theta_{23}} +\sqrt{r_{c}} 
         \right]\;.\iran
\ena
In addition, we have 
\bea
 \left.
  \bear{l}
    \det{H_{d}} = m_d^2m_s^2m_b^2=\lambda^{14} r_{d}^2 r_{s}^4 m_{b}^6\;,\\
    {\rm tr}{H_{d}} =  m_d^2+m_s^2+m_b^2=
    (1+\lambda^5 r_{s}^2 +\lambda^9 r_{s}^2
                    r_{d}^2)m_{b}^2\;.
  \enar
 \right.
\ena
 
Then, the elements of $\nami M_{d}$ are reconstructed as 
\bea
   a_{d} &\simeq& m_{b}\lambda^{7/2}\sqrt{ r_{s}^2 + \lambda
         f_{1}}\;,\iran \\
   b_{d} &\simeq& m_{b}\lambda^{7/2} r_{d}r_{s}^2 
              \sqrt{ \frac{1}{ r_{s}^2 -\lambda f_{2} } }\;,
              \iran\\
   c_{d} &\simeq& m_{b}\lambda^2 
          \sqrt{ \frac{\left( A^2 +r_{c} +f_{3} \right)
                       \left( r_{s}^2 + \lambda f_{1} \right) }
                      { r_{s}^2 -\lambda f_{2} } }\;,\iran \\
   d_{d} &\simeq& m_{b}\lambda^{1/2}
              \sqrt{ \frac{ \sigma^2 +A^2 r_{u} -f_{4} }
                          { r_{s}^2 + \lambda f_{1} } }\;,\\     
   e_{d} &\simeq& m_{b}\sqrt{ \frac{ r_{s}^2 -\lambda f_{2} }
                          { r_{s}^2 + \lambda f_{1}}
                          }\;,\iran\\
   \alpha_{1} &\simeq& 
          \theta_{23}+\tan^{-1}{ \left(
             \frac{ \sigma\sin{(\theta_{12}-\delta)} }
                  {\sigma\cos{(\theta_{12}-\delta)} -A\sqrt{r_{u}} }
                 \right) }\;, \iran \\
   \alpha_{2} &\simeq&
          \tan^{-1}{\left( \frac{ A \sin{\theta_{23}} }
                          { A \cos{\theta_{23}}+\sqrt{r_{c}} }
                     \right)}\; \iran 
\ena 
where 
\bea
 \left.
  \bear{l}
    f_{1} \equiv \sigma^2 +A^2 r_{u} -2\sqrt{r_{u}} 
                           ( A \sigma \cos{(\theta_{12}-\delta)}
                            +r_{s}^2 \cos{\theta_{12}})\;,\\  
    f_{2} \equiv 2 \sqrt{r_{u}}r_{s}^2 \cos{\theta_{12}}\;,\\ 
    f_{3} \equiv 2 A \sqrt{r_{c}} \cos{\theta_{23}}\;,\\
    f_{4} \equiv 2 A \sigma
        \sqrt{r_{u}}\cos{(\theta_{12}-\delta)}\;.
  \enar
 \right.
\ena 
These expressions contain phases $\theta_{12}$, $\theta_{12}-\delta$  
and $\theta_{23}$ which are fixed by solving Eq.(4) or (15). That is, 
from Eq.(17), $\cos (\theta_{23}-\zeta)$ is given and  then 
the angle $\theta_{23}$ is obtained as
\bea
 \left.
  \bear{l}
   \cos{\theta_{23}} =\left[ -Z(p+q\cos{\delta})
                            \pm\sqrt{4N^2-Z^2}q\sin{\delta}
                      \right]/2N^2\;,\\
   \sin{\theta_{23}} =-\left[\pm\sqrt{4N^2-Z^2}(p+q\cos{\delta})
                            +Zq\sin{\delta}
                      \right]/2N^2\;,
  \enar
 \right.
\ena
where the upper sign corresponds to positive sign case of 
$\sin{(\theta_{23}-\zeta)}$ and the lower sign does to negative sign. 
Next, by using $\exp({i(\theta_{12}-\delta)}) 
       =( D_{1} +B_{1} e^{i \theta_{23}} )/
         (|A_{1}|\exp({-i(\kappa-\delta)})+C_{1}\exp({i\theta_{23}}))$
we obtain 
\bea
 \cos\left(\theta_{12}-\delta \right)
   &=& \left[ B_{1}C_{1} +\lambda^6 D_{1}A\sigma 
             +D_{1}\lambda^6 r_{s}^2\cos{\delta}
             +B_{1}\lambda^6 r_{s}^2 \sin{\delta}\sin{\theta_{23}}
              \right.\iran\\ 
   & &\mbox{ } \left.+\left(C_{1}D_{1} +\lambda^6 B_{1}A\sigma 
                            +B_{1}\lambda^6 r_{s}^2 \cos{\delta} 
                      \right)\cos{\theta_{23}}
       \right]/M\;,\nonumber\\
 \sin \left(\theta_{12}-\delta\right) 
   &=& \left[ \left\{ \lambda^6 B_{1}(A\sigma +r_{s}^2\cos{\delta})
                     -C_{1}D_{1}
              \right\} \sin{\theta_{23}}\right. \iran\\
          &&\mbox{}\left. 
          -(D_{1}+B_{1}\cos{\theta_{23}})\lambda^6 r_{s}^2 \sin{\delta} 
       \right]/M\;,
\ena
where,
\bea
 \lefteqn{M} \quad
 &\equiv& (\lambda^6 A \sigma)^2 +(\lambda^6 r_{s}^2)^2 +C_{1}^2 
         +2 A \sigma \lambda^{12} r_{s}^2 \cos{\delta} \iran\\
 && \mbox{}+2\lambda^6 C_{1}\left\{ (r_{s}^2 \cos{\delta} +A \sigma)
                      \cos{\theta_{23}}
                   +r_{s}^2 \sin{\delta}\sin{\theta_{23}}
            \right\}\;.
\ena
Finally, $\theta_{12}$ is obtained by using Eq.(30). 

By substituting these angles, the mass matrix 
$M_{d}$ is expressed in terms of
quark masses and CKM parameters. 

 Since expressions   
are complicated, we show the result numerically. For quark masses and 
$\lambda$, we take the values given before and we use the central 
values for $A$ and $\sigma$, i.e., $|K_{cb}|$ and $|K_{ub}|$. Then, 
elements of $\nami M_d$ depend  on the CP violating parameter $\delta$. 
Thus, we present the values of elements as a function of $\delta$.  
We show the result in Fig.3 and Fig.4 which correspond 
to positive and negative values of 
$\sin{(\theta_{12}-\zeta)}$, respectively.  The region where the 
transformation is possible is $\cos{\delta} \leq 0.66$~\cite{Taka}.  

These figures show that there is no symmetric matrices, i.e., 
$|a_d|=b_d$ and $|c_d|=d_d$ in the NNI 
form which reproduce the present data. The constraint 
$|a_d|=b_d$ is possible for $\delta \sim 130^o$, $\sim 60^o$, 
but $|c_d|$ is different from $d_d$. 
In Fig.4, we see that $|a_{d}|$ approximately equals $b_{d}$, and
$d_{d}$ does $e_{d}$ for $\delta \sim 60^o$ which is the case 
where the Fritzsch $Ansatz$ for $M_u$ and the BS $Ansatz$ for $M_d$.

\section{Discussions}

We showed that the transformation of quark mass matrices 
which leaves the left-handed gauge interaction invariant  
to the Fritzsch-type 
form or to the BS-type form is possible in ``general". 
The word ``general" means that it is possible for almost all 
quark mass matrices which reproduce the present data of 
CKM mixings and the CP violating phase. Thus, we concluded 
that the above forms are most general forms which we can 
start in general. By using the Fritzsch-type form, we reconstructed 
quark mass matrices from quark masses, CKM mixings and the CP 
violating parameter. We calculated each elements of mass matrices 
numerically and showed in Figs.3 and 4. As it is already known, 
the symmetric mass matrices are not allowed in this parametrization 
so that the asymmetric form is unavoidable. 

In this paper, we discussed quark mass matrices at $M_Z$ scale, 
but it is more interesting to have them at the unification scale 
since they may be related directly to the structure of Yukawa 
interactions in grand unified theories.  
This can be done immediately by following our procedure once 
the data at the unification scale are given. Thus, it is 
important to obtain the CKM mixings at the unification scale 
from those at the weak scale, which is now under consideration. 
 
The asymmetric form of quark mass matrices is inevitable for the NNI 
basis so that we have to go to another type form  if we prefer to use  
symmetric form. The symmetric form arises naturally in  models like 
SO(10) unification.  As for the symmetric form, the most interesting 
choice will be the one where the 1-1 and 1-3 elements are zero.  
It should be possible to transform  quark masses in this form in 
general and  reconstruct the quark mass matrices from the data. This 
is  now under investigation.

Finally, from our analysis we can say what quark mass matrices are 
required once the CP violating phase is fixed  in the Fritzsch-type 
NNI parametrization.  

\newpage

\newpage
\noindent
Figure caption

\begin{itemize}
\renewcommand{\labelitemi}{--}
\item Fig.1 : The CKM parameter region where the transformation 
      to the Fritzsch-type parametrization is possible. 
      The allowed region is shown by the dotted area 
      which is almost independent of $A$. 
      These figures show that for most of physical quark mass 
      matrices the transformation is possible. 
      The F-BS point shows the predicted point 
      when the Fritzsch form is assumed 
      for $M_u$  and the BS form for $M_d$.
\item Fig.2 : The CKM parameter region where the transformation 
      to the BS-type parametrization is possible.  
      These figures show that for all physical quark mass 
      matrices the transformation is possible. 
\item Fig.3 : The reconstructed elements of $M_d$ 
      in the Fritzsch-type parametrization for 
      $\sin{(\theta_{12}-\zeta)}>0$. 
      In this case, $M_u$ is in the Fritzsch 
      form and is expressed by u-quark masses. 
\item Fig.4 : The reconstructed elements of $M_d$ 
      in the Fritzsch-type parametrization for 
      $\sin{(\theta_{12}-\zeta)}<0$. 
      The F-BS point corresponds to the $\delta \sim 60^o$ 
      where $|a_d|=b_d$ and $d_d \simeq e_d$. 
\end{itemize}    
 \begin{figure}[p]
 \begin{center}
  {\large Fig.1}
 \end{center}
 \epsfxsize=11.5cm
 \centerline{\epsfbox{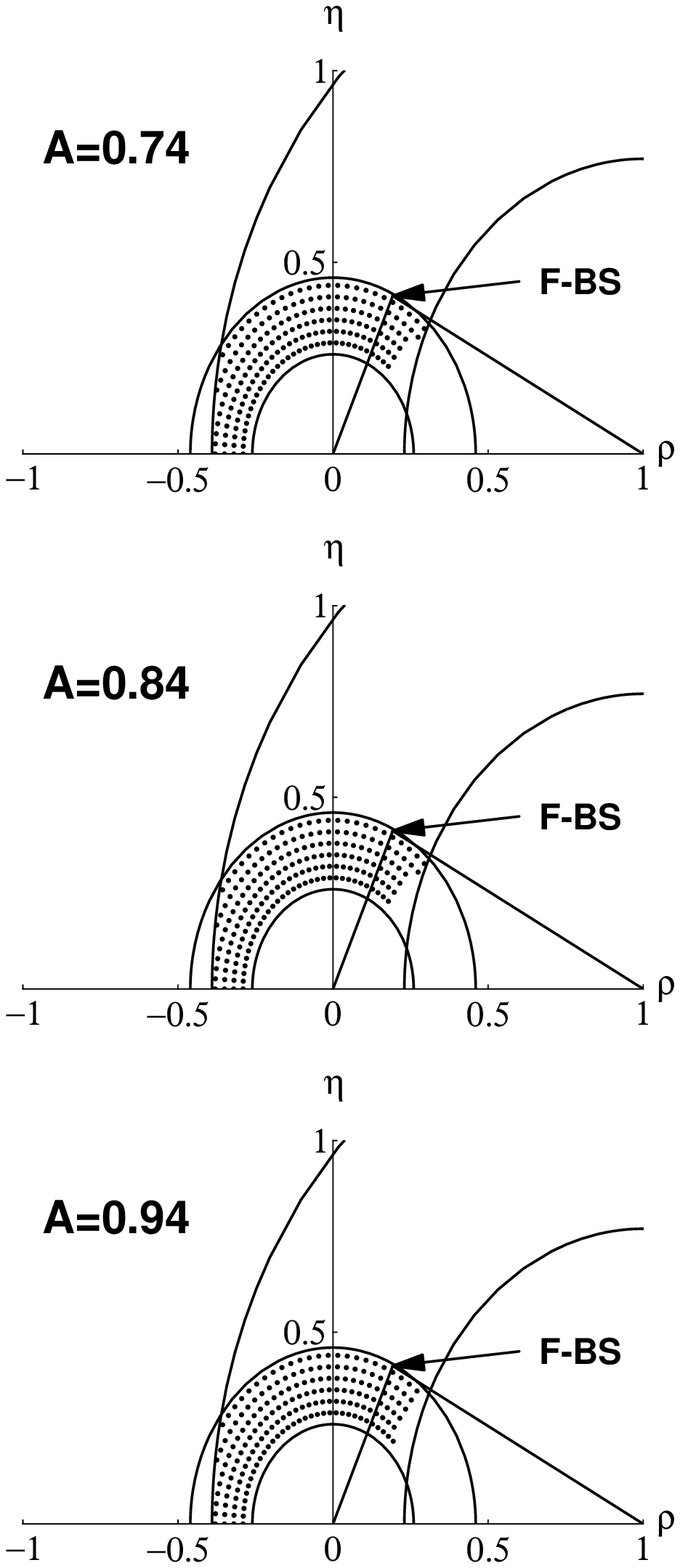}}
 \label{fig1}
 \end{figure}
 \begin{figure}[p]
 \begin{center}
  {\large Fig.2}
 \end{center}
 \epsfxsize=11.5cm
 \centerline{\epsfbox{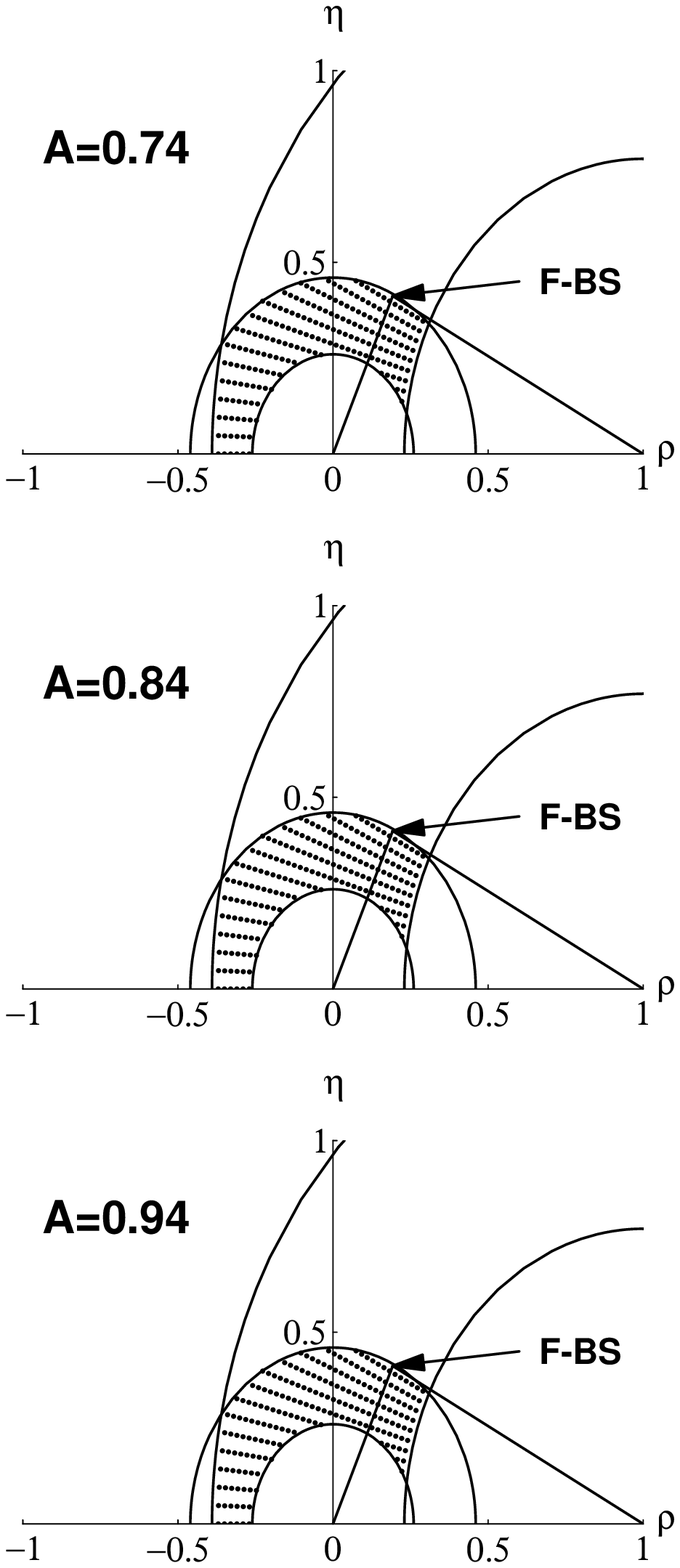}}
 \label{fig2}
 \end{figure}
 \begin{figure}[p]
 \begin{center}
  {\large Fig.3}
 \end{center}
 \epsfxsize=15.2cm
 \centerline{\epsfbox{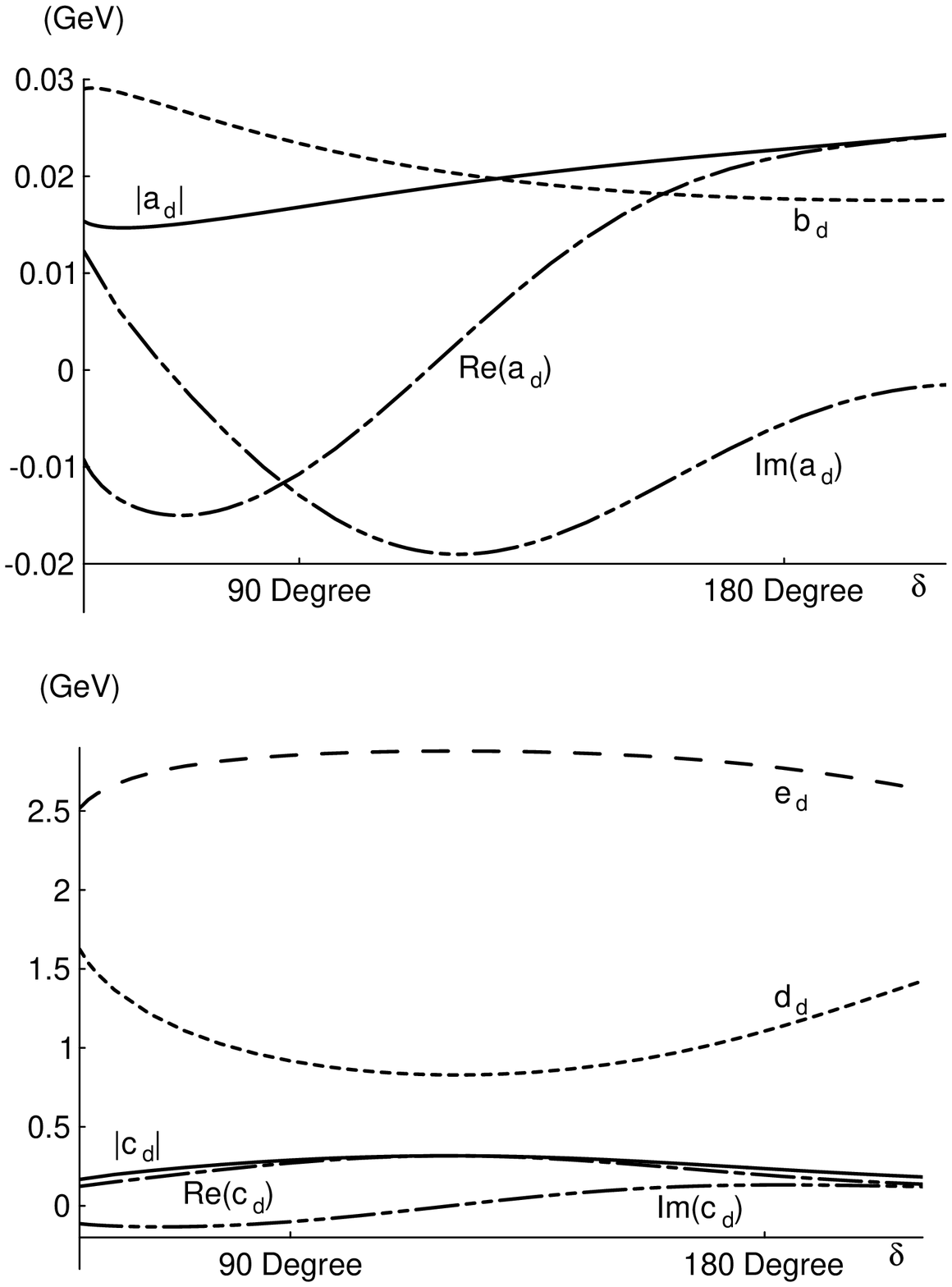}}
 \end{figure}
 \begin{figure}[p]
 \begin{center}
  {\large Fig.4}
 \end{center}
 \epsfxsize=15.2cm
 \centerline{\epsfbox{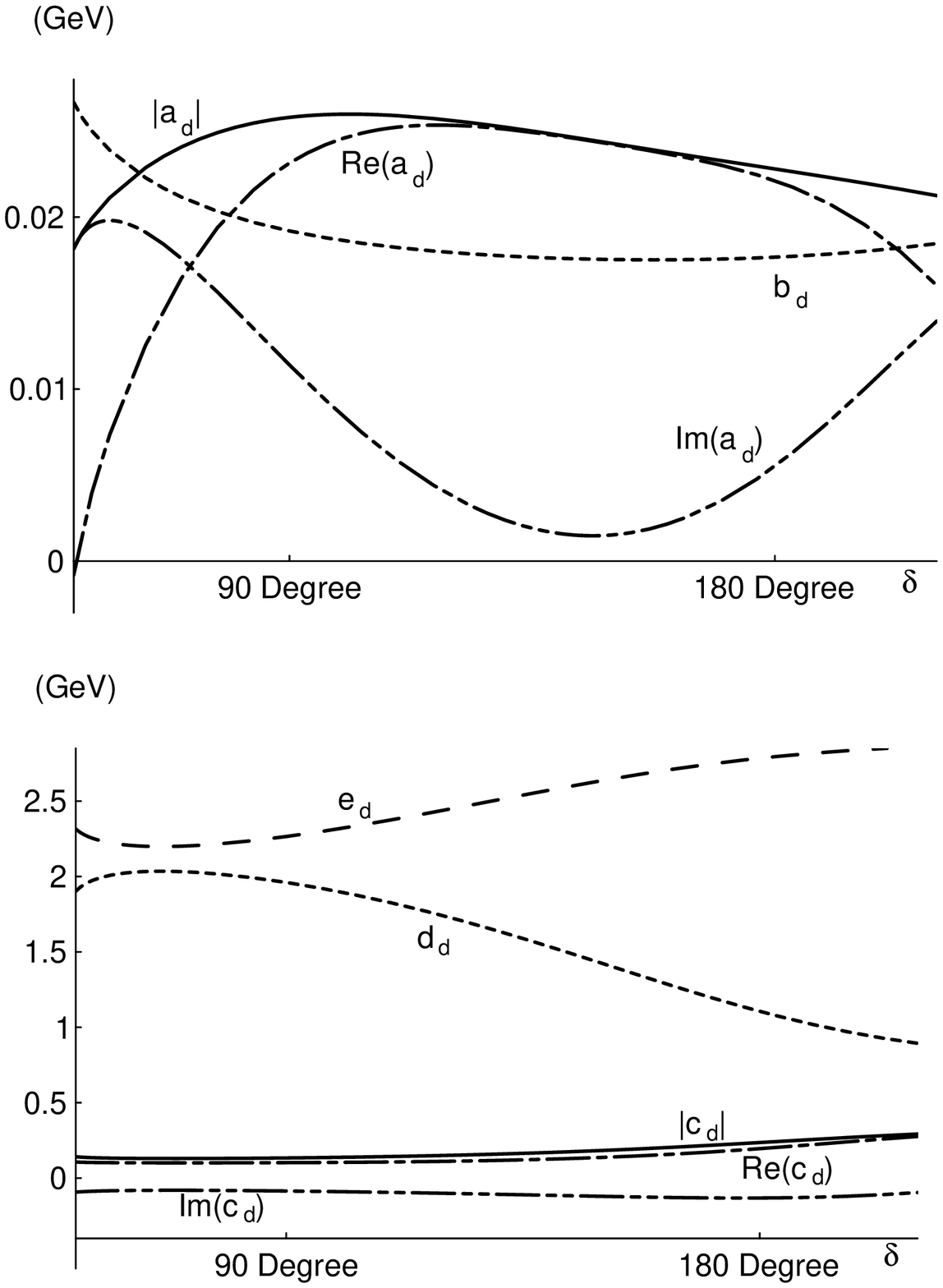}}
 \label{fig4}
 \end{figure}
\end{document}